\documentclass[preprint,showpacs,amsmath,superscriptaddress,nofootinbib
]{revtex4}

\usepackage{mathrsfs}
\usepackage{amssymb}
\usepackage{dcolumn}
\usepackage{bm}
\usepackage{epsfig}
\usepackage{hyperref}
\usepackage{graphicx}
\usepackage{dcolumn}
\usepackage{bm}
\usepackage{longtable}
\begin{document}

\title{Remarks on the two-dimensional power correction in the soft wall model}
\author{ZUO Fen} \footnote{Email: zuof@mail.ihep.ac.cn.}
\affiliation{Department of Modern Physics, University of Science and
Technology of China, Hefei, Anhui 230026, China}

\affiliation{Institute of High Energy Physics, Chinese Academy of
Sciences, Beijing 100049, China}

\author{HUANG Tao} \footnote{Email: huangtao@mail.ihep.ac.cn.}
\affiliation{Institute of High Energy Physics, Chinese Academy of
Sciences, Beijing 100049, China}

\begin{abstract}
We present a direct derivation of the two-point correlation function
of the vector current in the soft wall model by using the AdS/CFT
dictionary. The resulting correlator is exactly the same as the one
previously obtained from dispersion relation with the same spectral
function as in this model. The coefficient $C_2$ of the
two-dimensional power correction is found to be $C_2=-c/2$ with $c$
the slope of the Regge trajectory, rather than $C_2=-c/3$ derived
from the strategy of first quantized string theory. Taking the slope
of the $\rho$ trajectory $c\approx0.9\mbox{GeV}^2$ as input, we then
get $C_2\approx-0.45\mbox{GeV}^2$. The gluon condensate is found to
be $<\alpha_sG^2>\approx0.064\mbox{GeV}^4$, which is almost
identical to the QCD sum rule estimation. By comparing these two
equivalent derivation of the correlator of scalar glueball operator,
we demonstrate that the two-dimensional correction can't be
eliminated by including the non-leading solution in the
bulk-to-boundary propagator, as was done in \cite{Colangelo2}. In
other words, the two-dimensional correction does exist in the scalar
glueball case. Also it is manifest by using the dispersion relation
that the minus sign of gluon condensate and violation of the low
energy theorem are related to the subtraction scheme.
\end{abstract}

 \maketitle

\section{Introduction}
The standard operator product expansion (OPE) for the two point
correlator of the vector current $J_\mu=(\bar u\gamma_\mu u-\bar
d\gamma_\mu d)/\sqrt{2}$ has the following form:
\begin{equation}
\Pi_{\mu\nu}=i\int{d^4xe^{iq\cdot
x}<0|T\{J_\mu(x),J^\dagger_\nu(0)\}|0>=(q_\mu
q_\nu-q^2g_{\mu\nu})\Pi_V(Q^2=-q^2)},
\end{equation}
with
\begin{equation}
\mathcal{N}Q^2\frac{d\Pi(Q^2)}{dQ^2}=C_0+\frac{1}{Q^2}C_2+\sum_{n\ge2}\frac{n}{Q^{2n}}C_{2n}<\mathcal{O}_{2n}>,\label{eq:OPE1}
\end{equation}
where $C_0$ is given as $C_0=1+\Sigma_{n\ge1}A_n\alpha_s^n(Q^2)$,
and  $C_4< \mathcal{O}_4>=\frac{\pi}{3}<\alpha_sG^2>$ stands for the
gluon condensate. The two-dimensional correction $\frac{1}{Q^2}C_2$
does not appear in QCD with massless quarks \cite{SVZ}. However, it
was argued in Ref.\cite{Zakharov} that a non-vanishing $C_2$ may be
related to the string effect. It is also suggested that the origin
of these corrections is due to nonlocal properties of the instanton
vacuum \cite{Dorokhov}. In Ref.\cite{Shifman, Arriola}, it was found
that in the large-$N_c$ limit radial Regge trajectories naturally
give rise to the presence of the two-dimensional correction, unless
fine-tuning the parameters. Recently this term was found
\cite{Andreev} to appear in the so called soft wall model
\cite{Karch}, which can accomplish linear Regge trajectories very
well based on gauge/string duality \cite{Maldacena}. The coefficient
$C_2$ is derived to be given by the Regge slope $c$, $C_2=-c/3$,
following the strategy of first quantized string theory. On the
other hand, this correlator can be derived from the dispersion
relation since the masses and decay constants have be given in this
model. The result for $C_2$ seems to be a little different
\cite{Cata}. In order to determine the true correlator and
consequently the value of $C_2$, we give a direct derivation based
on the original AdS/CFT prescription. The resulting correlator is
exactly the same as the one derived from the dispersion relation,
showing the equivalence of these two approaches. The coefficient
$C_2$ of the two-dimensional correction is found to be $C_2=-c/2$,
which has the same sign as in QCD if quark massed are taken into
account. Taking the slope of the $\rho$ trajectory
$c\approx0.9\mbox{GeV}^2$ as input, we then get
$C_2\approx-0.45\mbox{GeV}^2$. The gluon condensate is also
calculated and the numerical result is given by
$<\alpha_sG^2>\approx0.064\mbox{GeV}^4$, which is almost identical
to the common QCD sum rule estimation \cite{Narison}.

The correlator of two-point scalar glueball operator in the same
model was discussed in \cite{Forkel, Colangelo2}. With the standard
choice for the bulk-to-boundary propagator, the correlator also
contains a similar two-dimensional power correction (times a
logarithm). Moreover, by expanding the correlator in large $Q^2$
limit and comparing to the OPE, one get a negative gluon condensate
contrast to the common value. It seems that by including another
solution for the bulk-to-boundary and fine-tuning the coefficients,
the two-dimensional correction can be eliminated and the common
value of the gluon condensate can be obtained \cite{Colangelo2}.
However, since the scalar glueball spectrum and the corresponding
residues have been given in this model, we can also get the
correlator by using the dispersion relation. Again the result
coincides with the one calculated in the standard way \cite{Forkel}.
Thus including of the other solution in the bulk-to-boundary
propagator is redundant. Actually,  with the normalizable modes in
hand, using of the decomposition formula \cite{Strassler}
immediately results the bulk-to-boundary propagator with the
standard choice, just as in the vector case \cite{Radyushkin}. From
the procedure of the dispersion relation it can also be found that
the determination of gluon condensate is substraction scheme
dependant.

\section{the two point vector correlator in the soft model}
The soft-wall model can be given by introducing a non-constant
dilaton field \cite{Karch} $\Phi(z)=\kappa^2z^2$ to the original AdS
metric
\begin{equation}
ds^2=e^{2A(z)}(-dz^2+\eta_{\mu\nu}dx^\mu~dx^\nu)
\end{equation}
with $A(z)=-\log z$. The relevant action for the vector field is
given by:
\begin{equation}
I=\int~d^5xe^{-\Phi(z)}\sqrt{g}\mbox{Tr}\left\{-\frac{1}{g_5^2}(F_L^2+F_R^2)\right\},
\end{equation}
where $A_{L,R}=A^a_{L,R}t^a, F_{MN}=\partial_M A_N-\partial_N
A_M-i[A_M,A_N]$ and $t^a=\sigma^a/2$. The vector current
$J_\mu=(\bar u\gamma_\mu u-\bar d\gamma_\mu d)/\sqrt{2}$ is
considered to be dual to the vector combination $V=(A_L+A_R)/2$. We
can use the gauge invariance of the action to go to the axial gauge
$V_z=0$, and derive the equation for the 4d-transverse components of
$V_\mu$. It is found to be
\begin{equation}
\left[\partial_z\left(\frac{e^{-\Phi}}{z}\partial_zV^a_\mu(q,z)\right)+\frac{e^{-\Phi}}{z}q^2V^a_\mu(q,z)\right]_\perp=0,\label{eq:vector}
\end{equation}
where $V^a_\mu(q,z)$ is the 4D Fourier transform of the original
field $V^a_\mu(x,z)$. Substituting the solution to
Eq.(\ref{eq:vector}) to the action, we get a surface term
\begin{equation}
I=-\frac{1}{2g_5^2}\int{d^4x\left(\frac{e^{-\Phi}}{z}V^a_\mu\partial_zV^{\mu
a}\right)}_{z=\epsilon}\label{eq:surface}.
\end{equation}
The non-normalizable solution of $V^a_\mu$ can be expressed by
$V_\mu(q,z)=V(q,z)V_\mu^0(q)$, where $V_\mu^0(q)$ is the Fourier
transform of the source coupled to $J_\mu$, and $V(q,z)$ is the
bulk-to-boundary propagator. $V(q,z)$ must satisfies the boundary
condition $V(q,z=0)=1$ to ensure the interpretation of $V^\mu_0(q)$
as the source. According to the AdS/CFT dictionary \cite{Polyakov,
Witten}, the partition function on the AdS side is equal to the
generation function of the dual CFT. Thus we can obtain the
two-point function by differentiating Eq.(\ref{eq:surface}) twice
with respect to the source $V_0$
\begin{equation}
\Pi_V(Q^2)=-\frac{1}{g_5^2Q^2}
\left[\frac{e^{-\Phi}}{z}V(Q,z)\partial_zV(Q,z)\right]_{z=\epsilon}
\end{equation}
with $Q^2=-q^2$. The general solution of Eq.(\ref{eq:vector}) is
given by the confluent hypergeometric functions
\begin{equation}
V(Q,z)=A_V~U(\frac{Q^2}{4\lambda^2},0,\lambda^2z^2)+B_V~\lambda^2z^2~_1F_1(\frac{Q^2}{4\lambda^2}+1,2,\lambda^2z^2).
\end{equation}
Notice that in this case we should choose the regular solution
$\lambda^2z^2~_1F_1(\frac{Q^2}{4\lambda^2}+1,2,\lambda^2z^2)$,
rather than $_1F_1(\frac{Q^2}{4\lambda^2},0,\lambda^2z^2)$
\cite{Radyushkin}, since the latter is not well-defined
mathmatically \cite{Erdelyi}. Recently there is some controversy in
choosing the solution of $V(Q,z)$ \cite{Forkel,Colangelo2}, which we
will discuss in the next section. Now we just follow the standard
procedure and set $B_V=0$. Then $A_V$ can be determined from the
boundary condition to be $A_V=\Gamma(1+Q^2/4\lambda^2)$. Thus the
correlator is given by
\begin{equation}
\Pi_V(Q^2)=-\frac{1}{g_5^2Q^2}\Gamma\left(\frac{Q^2}{4\lambda^2}+1\right)\lim_{z\to
0}\frac{e^{-\Phi}}{z}\partial_zU\left(\frac{Q^2}{4\lambda^2},0,\lambda^2z^2\right)\label{eq:VC3}
\end{equation}
Using the integral representation of the Tricomi function
\cite{Erdelyi}
\begin{equation}
U(a,c;x)=\frac{1}{\Gamma(a)}\int^1_0du\frac{u^{a-1}}{(1-u)^c}\exp\left[-\frac{u}{1-u}x\right]
\end{equation}
Eq.(\ref{eq:VC3}) can be simplified to be
\begin{equation}
\Pi(Q^2)=\frac{1}{2g_5^2}\Gamma\left(\frac{Q^2}{4\lambda^2}+1\right)
\lim_{z\to
0}U\left(\frac{Q^2}{4\lambda^2}+1,1,\lambda^2z^2\right).\label{eq:VC4}
\end{equation}
The Tricomi function can be expanded as \cite{Erdelyi}
\begin{eqnarray}
U(a,n+1;x)=&&\frac{(-1)^{n-1}}{n!\Gamma(a-n)}\{_1F_1(a,n+1;x)\log~x\\\nonumber
~~~~~~~~~~~&&+\sum_{r=0}^\infty\frac{(a)_r}{(n+1)_r}[\psi(a+r)-\psi(1+r)-\psi(1+n+r)]\frac{x^r}{r!}\}\\\nonumber
~~~~~~~~~~~&&+\frac{(n-1)!}{\Gamma(a)}\sum_{r=0}^{n-1}\frac{(a-n)_r}{(1-n)_r}\frac{x^{r-n}}{r!}~~~~~n=0,1,2,...,
\end{eqnarray}
where $(\lambda)_n=\Gamma(\lambda+n)/\Gamma(\lambda)(n\ge1),
(\lambda)_0=1$ and $\psi(x)$ is the digamma function defined as
\begin{equation}
\psi(x)=\frac{d}{dx}\log{\Gamma(x)}.
\end{equation}
Taking the $z\to 0$ limit in Eq.~(\ref{eq:VC4}) we can easily get
the expression for the correlator
\begin{equation}
\Pi_V(Q^2)=-\frac{1}{2g_5^2}\psi\left(1+\frac{Q^2}{4\lambda^2}\right)\label{eq:VC1}
\end{equation}
up to one irrelevant (subtraction) constant which is actually
infinite. This kind of correlator has been discussed extensively in
the literature \cite{Shifman,Cata}, obtained from dispersion
relation with a model for the spectral function. More specifically,
the correlation function can be expressed by the spectral function
by using the dispersion relation
\begin{equation}
\Pi_V(q^2)=\frac{1}{\pi}\int^\infty_0\frac{\mbox{Im}\Pi(s)}{s-q^2}ds\label{eq:DR}
\end{equation}
In the large-$N_c$ limit, the spectral function is given by an
infinite number of resonances
\begin{equation}
\frac{1}{\pi}\mbox{Im}\Pi(s)=\sum^\infty_{n=0}F_n^2\delta(s-m_n^2).\label{eq:SF}
\end{equation}
Furthermore, one assumes that the resonances lie on linear Regge
trajectory
\begin{equation}
m_n^2=m_0^2+cn,
\end{equation}
and the residues are independent of the excitation number $n$
\begin{equation}
F_n^2=F^2.
\end{equation}
With these relations the correlator can be calculated explicitly,
and is given by
\begin{equation}
\Pi_V(Q^2)=-\frac{F^2}{c}\left[\psi(\frac{Q^2+m_0^2}{c})+\mbox{Const}\right]\label{eq:VC2}
\end{equation}
Notice that in the soft wall model, Regge trajectory can be
reproduced naturally \cite{Karch}
\begin{equation}
m_n^2=4\lambda^2(n+1),
\end{equation}
and the residues are also constant
\begin{equation}
F^2=\frac{2\lambda^2}{g_5^2}.
\end{equation}
From these relations one can show that the correlator given in
Eq.(\ref{eq:VC1}) and Eq.(\ref{eq:VC2}) are exactly the same up to
the subtracted term. This in turn confirms our choice for the
bulk-to-boundary propagator.

The OPE of the correlator (\ref{eq:VC2}) can be obtained by using
the properties of the digamma function~\cite{Peris,Cata}
\begin{equation}
\Pi^{\rm{OPE}}_V(Q^2)=-\frac{F^2}{c}\log{\frac{Q^2}{\mu^2}}+\sum_{k=1}^\infty\frac{c_{2k}}{Q^{2k}},
\end{equation}
where the condensates are given by
\begin{equation}
c_{2k}=(-1)^k\frac{F^2c^{k-1}}{k}B_k(\frac{m_0^2}{c}),
\end{equation}
and $B_k(\xi)$ denotes the Berboulli polynomials. In our simpler
case (\ref{eq:VC2}) we can get the OPE directly since
\begin{equation}
\psi(\xi+1)=\psi(\xi)+\frac{1}{\xi},
\end{equation}
and the large-$\xi$ behavior of the digamma function is
\begin{equation}
\psi(\xi)=\log{\xi}-\frac{1}{2\xi}-\sum_{n=1}^\infty\frac{B_{2n}}{2n\xi^{2n}}
\end{equation}
with the Bernoulli numbers
$B_{2n}=(-1)^{n-1}2(2n)!\zeta(2n)/(2\pi)^{2n}$ ($\zeta(z)$ is
Riemann's zeta function). The result becomes
\begin{equation}
\Pi^{\rm{OPE}}_V(Q^2)=-\frac{1}{2g_5^2}\log{\frac{Q^2}{4\lambda^2}}-\frac{\lambda^2}{g_5^2Q^2}+\frac{1}{2g_5^2}\sum_{n=1}^{\infty}\frac{B_{2n}(4\lambda^2)^{2n}}{2n(Q^2)^{2n}}.\label{eq:OPE2}
\end{equation}
Notice that the leading behavior is the same as that of the hard
wall model \cite{Erlich}. It shows that different deformation in the
IR region do not affect the large $Q^2$ behavior, which is governed
by the essentially AdS metric in the UV. This was also true in the
case of the scalar glueball \cite{Forkel}. Matching to the OPE in
real QCD to leading order in $\alpha_s$
\begin{equation}
\Pi^{\rm{OPE}}_V(Q^2)=-
\frac{2}{3}\frac{N_c}{(4\pi)^2}\log{\frac{Q^2}{\mu^2}}+...,
\end{equation}
we get the value $g_5^2=12\pi^2/N_c$ as in Ref. \cite{Erlich}. Now
applying the operator
\begin{equation}
D(Q^2)=-\frac{12\pi^2}{N_c}~Q^2\frac{d}{dQ^2}
\end{equation}
on both sides of
Eq.(\ref{eq:OPE2}) and comparing with Eq.(\ref{eq:OPE1})  we get
\begin{equation}
C_0=1,  ~~C_2=-2\lambda^2=-c/2,~~<\alpha_sG^2>=\frac{1}{4\pi}c^2.
\end{equation}
This is in contradiction with the result $C_2=-c/3$ derived in
Ref.~\cite{Andreev} by first quantizing string theory in the same
background. From the coincidence of the correlator derived from the
two different approaches, it seems that $C_2=-c/2$ is the right
result. Then we get the value $C_2\approx-0.45\mbox{GeV}^2$ if we
take the slope of the $\rho$ trajectory $c\approx0.9\mbox{GeV}^2$ as
input. The gluon condensate can also be obtained, which is given by
\begin{equation}
<\alpha_sG^2>\approx0.064\mbox{GeV}^4.
\end{equation}
Surprisingly, this result is almost identical to the QCD sum rule
estimation \cite{Narison}
\begin{equation}
<\alpha_sG^2>=(.06\pm.02)\mbox{GeV}^4.
\end{equation}
\section{Extending to the scalar glueball case}
Recently the scalar glueball in the same model has been discussed in
Ref.~\cite{Colangelo1, Forkel, Colangelo2}. The scalar glueball
operator $\mathcal{O}_{S}\left( x\right) =G_{\mu \nu }^{a}\left(
x\right) G^{a,\mu \nu }\left( x\right)$, which has dimension
$\Delta=4$, is considered to be dual to a massless scalar in the 5D
setup. The action is given by
\begin{equation}
I =\frac{1}{2\kappa ^{2}}\int d^{5}x\sqrt{ g }e^{-\Phi(z)
}g^{MN}\partial _{M}\varphi \partial _{N}\varphi
\end{equation}
The equation of motion for the Fourier transform $\varphi(q,z)$
reads
\begin{equation}
\left[\partial_z\left(\frac{e^{-\Phi}}{z^3}\partial_z\varphi(q,z)\right)+\frac{e^{-\Phi}}{z^3}q^2\varphi(q,z)\right]_\perp=0,\label{eq:glueball}
\end{equation}
For $q^2=4\lambda^2(n+2)$, the solutions are
normalizable\cite{Colangelo1, Colangelo2}
\begin{equation}
\varphi_n(z)=A_n\lambda^4z^4~_1F_1(-n,3;\lambda^2z^2)\label{eq:NS2}
\end{equation}
with $A_n^2=(n+1)(n+2)/2\lambda$\footnote{Notice in
Ref.\cite{Colangelo2} the factor $\lambda$ is missing in $A_n^2$ due
to an unusual normalization.}. From the AdS/CFT dictionary, these
give the spectrum of the scalar glueball $m_n^2=4\lambda^2(n+2)$.
The general non-normalizable solution to Eq.(\ref{eq:glueball}) is
given by
\begin{equation}
\varphi(Q,z)=A_G~U\left(\frac{Q^2}{4\lambda^2},-1;\lambda^2z^2\right)+B_G~\lambda^4z^4~_1F_1\left(\frac{Q^2}{4\lambda^2}+2,3;\lambda^2z^2\right)\label{eq:propagator2}
\end{equation}
with $Q^2=-q^2$.

Now we again have to choose the right solution. In
Ref.~\cite{Forkel} $B_G$ is simply set to be zero as in the standard
way, and the corresponding correlator was obtained
\begin{eqnarray}
\Pi_G\left(Q^{2}\right)&=&i\int\frac{d^{4}q}{\left(2\pi\right)^{4}}e^{-iq\cdot
x}\left\langle~T\mathcal{O}_{S}\left(x\right)\mathcal{O}_{S}\left(0\right)\right\rangle\nonumber\\
                           &=&-\frac{\lambda ^{4}}{\kappa ^{2}}\frac{%
Q^{2}}{4\lambda ^{2}}\left( \frac{Q^{2}}{4\lambda ^{2}}+1\right)
\Gamma \left( \frac{Q^{2}}{4\lambda ^{2}}+1\right) \lim_{\varepsilon
\rightarrow 0}U\left( \frac{Q^{2}}{4\lambda ^{2}}+2,2,\lambda
^{2}\varepsilon ^{2}\right) \\
                    &=&=-\frac{2\lambda ^{4}}{\kappa ^{2}}\left[
1+\frac{Q^{2}}{4\lambda ^{2}}\left( 1+\frac{Q^{2}}{4\lambda
^{2}}\right) \psi \left( \frac{Q^{2}}{4\lambda ^{2}}\right)
\right]\label{eq:Pi0}\\
                    &=&-\frac{1}{8\kappa ^{2}}Q^{4}\left[ \ln \frac{Q^{2}}{\mu
^{2}}+\frac{4\lambda
^{2}}{Q^{2}}\ln \frac{Q^{2}}{\mu ^{2}}+\frac{2^{2}5}{3}\frac{\lambda ^{4}}{%
Q^{4}}+...\right] .
\end{eqnarray}%
where two contact terms proportional to $Q^2$ and $Q^4$ have been
discarded. The corresponding OPE in QCD reads
\begin{equation}
\Pi_G^{\left( \text{OPE}\right) }(Q^{2}) =A_{0} Q^{4}\ln \left(
\frac{Q^{2}}{\mu ^{2}}\right)+B_{0}\left\langle G^{2}\right\rangle
+O\left(\frac{1}{Q^2}\right)
\end{equation}
with $A_{0}=-\left( N_{c}^{2}-1\right) /\left( 4\pi ^{2}\right) $
and $B_{0}=4+49\alpha _{s}/\left( 3\pi \right) $ (for the number of
colors (flavors) $N_{c}$ $\left( N_{f}\right) =3$). As in the vector
case, a two-dimensional correction appeared here, just as in the
vector case. Also the gluon condensate can be obtained
\begin{equation}
\left\langle G^{2}\right\rangle  \simeq -\frac{10}{3\pi ^{2}}\lambda
^{4}, \label{g2}
\end{equation}
which is of the opposite sign to the common value $\left\langle
G^{2}\right\rangle \sim 0.4-1.2$ GeV$^{4}$. Actually, from
Eq.(\ref{eq:Pi0}) it can be seen that $\hat\Pi(0)=0$, thus a
non-vanishing value of $\left\langle G^{2}\right\rangle$ imply the
violation of the low energy theorem
\begin{equation}
\Pi_G\left( 0\right) =\frac{32\pi }{\alpha _{s}b_{0}}\left\langle
G^{2}\right\rangle +O\left( m_{q}\right). \label{let}
\end{equation}%

Later in Ref.\cite{Colangelo2} the full solution
(\ref{eq:propagator2}) was considered, with $B_G$ non-vanishing and
as a function of $Q^2$. By fine-tuning the expansion coefficients of
$B_G$ in the large $Q^2$ region, it was shown that the
two-dimensional correction can be eliminated. Moreover, a acceptable
value of $\left\langle G^{2}\right\rangle$ can be obtained and the
low energy theorem can be resumed.

To clarify this issue, we can calculate the correlator using the
dispersion relation. For this purpose we need the residues
\begin{equation}
F_n\equiv\left\langle0\left|\mathcal{O}_{S}(0)\right|n\right\rangle.
\end{equation}
This can be easily calculated from the normalizable solution
(\ref{eq:NS2}) and was given by \cite{Forkel,Colangelo2}
\begin{equation}
F_n^2=8(n+1)(n+2)\lambda^6/\kappa^2.\label{eq:DC}
\end{equation}
Then we can get the spectral function using the definition
(\ref{eq:SF})
\begin{equation}
\frac{1}{\pi}\mbox{Im}\Pi_G(s)=\frac{8\lambda^6}{\kappa^2}\sum_{n=0}^\infty
(n+1)(n+2)\delta \left( s-m_{n}^{2}\right)\label{eq:SF2}
\end{equation}
The behavior of $F_n^2$ as $n\to \infty$ suggests that we start from
the subtracted correclator
\begin{eqnarray}
\hat\Pi_G(q^2)&=&\Pi_G(q^2)-\Pi_G(0)-q^2\Pi_G'(0)-\frac{q^2}{2}\Pi_G^{''}(0)\nonumber\\
&=&\frac{q^6}{\pi}\int_0^\infty\frac{\mbox{Im}\hat\Pi(s)}{s^3(s-q^2)}
\end{eqnarray}
Now using the dispersion relation  (\ref{eq:DR}) we can find
\begin{eqnarray}
\hat\Pi_G(Q^2)&=&-Q^6\sum_{n=0}^\infty\frac{F_n^2}{m_n^6(m_n^2+Q^2)}\nonumber\\
              &=&-\frac{2\lambda^4}{\kappa^2}\left[1+\frac{Q^2}{4\lambda^2}\left(1+\frac{Q^2}{4\lambda^2}\right)\psi\left(\frac{Q^2}{4\lambda^2}\right)+\eta_1\frac{Q^2}{4\lambda^2}+\eta_2\left(\frac{Q^2}{4\lambda^2}\right)^2\right]
\end{eqnarray}
with $\eta_1=2-\psi(2)$ and
\begin{equation}
\eta_2=-\psi(2)-\sum_{n=2}^\infty\frac{1}{n^2}.
\end{equation}
Thus we explicitly reproduce the correlator (\ref{eq:Pi0}), up to
the subtracted terms. Actually, starting from the correlator
(\ref{eq:Pi0}), the spectral function has been obtained
\cite{Forkel} by using the causal pole definition
\begin{equation}
\frac{1}{\pi}\mbox{Im}\Pi_G(s)=\frac{\lambda ^{2}}{2\kappa
^{2}}s\left( s-m_{0}^{2}/2\right) \sum_{n=0}^{\infty }\delta \left(
s-m_{n}^{2}\right). \label{swsd}
\end{equation}%
which is in fact the same as Eq.(\ref{eq:SF2}).

The coincidence of the correlator calculated from two different
approaches again tells us that we should discard the non-leading
solution in the bulk-to-boundary propagator. This can be seen
directly from another way, using the decomposition formula derived
in Ref.\cite{Strassler}
\begin{equation}
\varphi(q,z)=\kappa\sum_n\frac{F_n\varphi_n(z)}{-q^2+m_n^2}
\end{equation}
where the factor $\kappa$ is added since in Ref.\cite{Strassler} the
"canonical normalization" $\varphi(q,0)=1/\kappa$ has been used.
Starting from the normalizable solution (\ref{eq:NS2}), we then
obtain the non-normalizable solution immediately
\begin{eqnarray}
\varphi(q,z)&=&\lambda^4z^4\sum_{n=0}^\infty\frac{L_n^2(\lambda^2z^2)}{a+n+2}\\
            &=&\Gamma(a+2)~U(a,-1;\lambda^2z^2)\label{eq:PG2}
\end{eqnarray}
where $a=-q^2/4\lambda^2$ and $L^\mu_n$ is the associated Lagurre
polynomials. In deriving the last expression we have used the
generation function of the Lagurre polynomial \cite{WZX}
\begin{equation}
\frac{e^{-\frac{zt}{1-t}}}{(1-t)^{\mu+1}}=\sum_{n=0}^\infty
L_n^\mu(z)t^n~~~(|t|<1).
\end{equation}
It can be confirmed that Eq.(\ref{eq:PG2}) gives the right boundary
normalization. Similar representation in the vector case has been
obtained in Ref.\cite{Radyushkin}.

Therefore the two-dimensional correction can't be eliminated by just
including the non-leading part in the bulk-to-boundary propagator.
Moreover, in our procedure an additional constant subtraction term
$\Pi_G(0)$ is needed but in Eq.(\ref{eq:Pi0}) this term is
identically zero. This confirms the conclusion in Ref.\cite{Forkel}
that the determination of the gluon condensate, and correspondingly
the low energy theorem, is dependant on the substraction procedure.

\section{summary}
The two-point function of the vector operator in the soft wall model
is derived by using the original AdS/CFT correspondence, which
coincides with that derived from the dispersion relation. A
non-vanishing two-dimensional correction is found in the OPE of this
correlator, with the coefficient $C_2$ different from the result
obtained from the strategy of first quantized string theory. The
numerical estimation of the gluon condensate is found to be almost
identical to the common value. From the equivalence of the two
derivations of the correlator, we conclude that the non-leading part
of the bulk-to-boundary propagator should be discarded.

Extending to the correlator of the scalar glueball operator, we
shows that the two-dimensional correction can not be eliminated by
just including the non-leading part. The odd sign of the gluon
condensate and the violation of the low energy theorem is found to
be related to the subtraction scheme.


\begin{thebibliography}{99}
\bibitem{Colangelo2}Colangelo P, De Fazio F, Jugeau F, and
Nicotri S arXiv: 0711.4747.

\bibitem{SVZ}Shifman M A, Vainshtein A I and Zakharov V I 1979 {\it Nucl. Phys. } {\bf B147}
385.

\bibitem{Zakharov}Zakharov V I 1998 {\it Prog. Theor. Phys. Suppl.}
{\bf {131}} 107.

\bibitem{Dorokhov}Dorokhov A E and Broniowski W 2003 {\it Eur. Phys. J. C.}
{\bf {32}} 79.

\bibitem{Shifman}Shifman M arXiv: hep-ph/0009131.

\bibitem{Arriola}Arriola E R and Broniowski W 2006{\it Phys. Rev. } {\bf D73}
097502.

\bibitem{Andreev}Andreev O 2006 {\it Phys. Rev. } {\bf D73}
107901.

\bibitem{Karch}Karch A, Katz E, Son D T, and Stephanov M A 2006 {\it Phys. Rev. } {\bf D74}
015005.

\bibitem{Maldacena}Maldacena J M 1998 {\it Adv. Theor. Math. Phys. }
{\bf 2} 231.

\bibitem{Cata}Cat\`{a} O 2007 {\it Phys. Rev. } {\bf D75}
106004.

\bibitem{Narison}Narison S 1989 {\it World Scientific Notes in Physics} vol
26.

\bibitem{Forkel}Forkel H arXiv: 0711.1179.

\bibitem{Strassler}Hoon S, Yoon S and Strassler M 2006 {\it JHEP } {\bf 04}
003.

\bibitem{Radyushkin}Grigoryan H R and Radyushkin A V 2007
{\it Phys. Rev. } {\bf D76} 095007.

\bibitem{Polyakov}
   Gubser S S, Klebanov I R and Polyakov A M 1998 {\it Phys. Lett. } {\bf B428} 105.

\bibitem{Witten}Witten E 1998 {\it Adv. Theor. Math. Phys. } {\bf 2} 253.

\bibitem{Erdelyi}Erd\'{e}lyi A 1953 {\it Higher Transcendental
Functions} Vol. I (New York-Toronto-London).

\bibitem{Peris}Cat\`{a} O, Golterman M and Peris S 2005 {\it JHEP } {\bf 08}
076.

\bibitem{Erlich}Erlich J, Katz E, Son D T, and Stephanov M A 2005 {\it Phys. Rev. Lett. } {\bf 95}
261602.

\bibitem{Colangelo1}Colangelo P, De Fazio F, Jugeau F, and
Nicotri S 2007 {\it Phys. Lett. } {\bf 652} 73.

\bibitem{WZX}Wang Z X and Guo D R 1989 {\it Special Functions}
(World Scientific Singapore).

























\end{thebibliography}
\end{document}